\documentclass[a4paper,11pt]{article}
\usepackage{pos}
\usepackage{multirow}

\title{Accretion Induced Collapse of White Dwarfs as an Alternative Symbiotic Channel to
Millisecond Pulsars}

\author*{Ali\, Taani}

\affiliation{Physics Department, Faculty of Science, Al Balqa Applied University, 19117 Salt, Jordan}

\emailAdd{ali.taani@bau.edu.jo}

\abstract{
Recently, extra motivation has been given to the investigations of an unresolved problem of millisecond pulsars (MSPs) produced by the recycling process, as an apparent role of the accretion-induced collapse (AIC) in  white dwarfs (WDs) was suggested to this concern. I have found that the distribution of the orbital periods of binary MSPs in the Galactic disk ($N_{obs,orb}$) closely follows an exponential distribution. I have also determined the best-fit mean value of $N_{obs,orb}$ by fitting our data with an exponential distribution for the MSP population.  As a result, it can be stated that reaching the Chandrasekhar limit  may  cause an explosion of a massive WD as a Type Ia supernova (in the case of a CO WD) or an ignition of a ONeMg WD, and possibly merging in some CO WDs, all  resulting in peculiar MSP systems. A possible formation scenario, where the system has a circular orbit during this evolutionary stage, is discussed.

}

\FullConference{%
  The Multifaceted Universe: Theory and Observations -  2022 (MUTO2022)\\
  23-27 May 2022\\
  SAO RAS, Nizhny Arkhyz, Russia\\}

\begin{document}
\maketitle

\section{Introduction}

Millisecond pulsars (MSPs) are remarkable objects born via Type II supernovae (SNe)
explosions, they are characterized by short spin periods (P$_{spin} \leq $ 20 ms), weak magnetic fields (B~$\leq 10^{9}$~G), and extremely old ages
$ \sim 10^{9}$ yr, based on the recycling process. The ATNF catalog counts 500 Galactic MSPs both in the Galactic plane and in the globular clusters \cite{2005AJ....129.1993M, 2008AIPC..968..194F}. They are often found in binaries (about 53\%) with white dwarfs (WDs) in circular
orbits, having companions with masses of $\rm \sim 0.15 M_{\odot} -  0.45M_{\odot}$ (see e.g., \cite{1982Natur.300..728A, 1991PhR...203....1B, 2022JHEAp..35...83T}).



It has been suggested that MSPs form in low-mass X-ray
binary systems.  This  argument leads to the recycling
process, in which a slowly rotating old neutron star (NS) may be spun up into an
MSP via accretion from a binary companion  \cite{1982Natur.300..728A, 2016RAA....16..101T, 2019AN....340..847T}.


An alternative formation process is an accretion-induced collapse (AIC) of
an ONeMg white dwarf (see, e.g., \cite{1987ApJ...322..206N, 2012AN....333...53T,2012Ap&SS.340..147T, 2017JPhCS.869a2090T,2017ApJ...846..170T, 2022MNRAS.513.4802A,2022MNRAS.tmp.2594M}). In this scenario,  the mass transfer can
increase the mass of the WD up to the Chandrasekhar limit.
This causes its collapse and a violent release of
gravitational energy, which might be observable by gravitational-wave observatories such as LIGO, VIRGO, and GEO~\cite{2010MNRAS.409..846D}.


\begin{figure*}

\begin{center}
\begin{tabular}{cc}
\includegraphics[width =15cm]{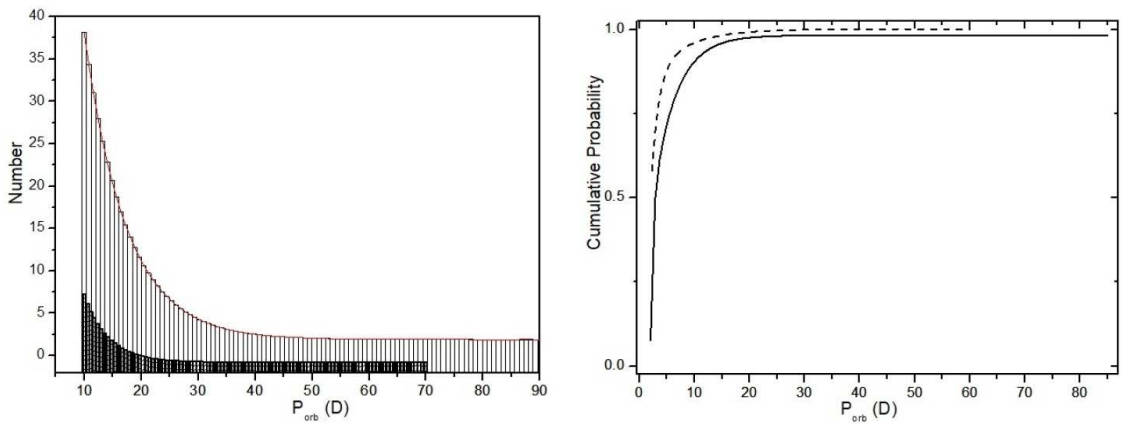}\\
\end{tabular}
\end{center}
\caption{Left: a histogram of the orbital periods of binary MSPs in the Galactic disk (white) and in the globular clusters (black). The solid line denotes the best fit to the data. Right: the cumulative probability distribution of the observed data for the Galactic disk (solid line) and the globular clusters (dashed line). The data were taken from the ATNF catalog (\cite{2005AJ....129.1993M}  and the Paulo Freire’s GC Pulsar Catalog (www.naic.edu/~pfreire/GCpsr.html).}
\label{fig2}
\end{figure*}

\begin{figure}
\begin{center}
\includegraphics[width=8.0cm, angle=0] {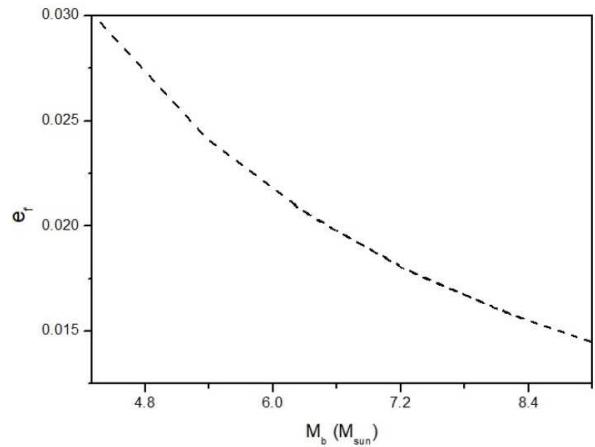}
\caption{
Final eccentricity $\rm e_{f}$ as a function of $M_{b}$.
}
\end{center}
\end{figure}

It is worth noting that in contrast to the Galactic field, globular clusters may provide  distinct  routes to the formation and evolution of accreting binaries.
As a result, the tidal capture process \cite{2007AIPC..921..249F, 2020arXiv200203011T} could be a source of long orbital period systems.
In addition, there is a possibility of the formation of such long-period binaries through mass exchange or mergers \cite{2017A&A...606A..45D, 2019ApJ...882...27M, 2019ApJ...875...89M,2022ApJ...936...78M}.
More observations in globular
clusters showed that some fraction of NSs must have received low kicks at birth. Fig. 1 shows the distribution of some representative observed data of the orbital periods of binary MSP systems in the Galactic disk and globular clusters (left) and  their  cumulative  probability  distribution (right). These parameters could play a role in examining the theoretical scenarios of the formation and evolution of recycled MSPs.
The  cumulative  probability  distribution  is best described by a log-normal law with  $\sigma =0.33$, $R^{2} =92\%$,  and $\chi^{2} =0.98$. In Fig. 2, we can see that the eccentricity $e$ is inversely proportional to the binary system mass $M_{b}$. It turns out that the low eccentricity is significant for the solutions presented for each system, as the circular orbit approximation is used. This generally reflects a substantial spread in orbits, as in Figs. 2 \& 3.

\begin{figure}
\begin{center}
\includegraphics[angle=0,width=8.0cm]{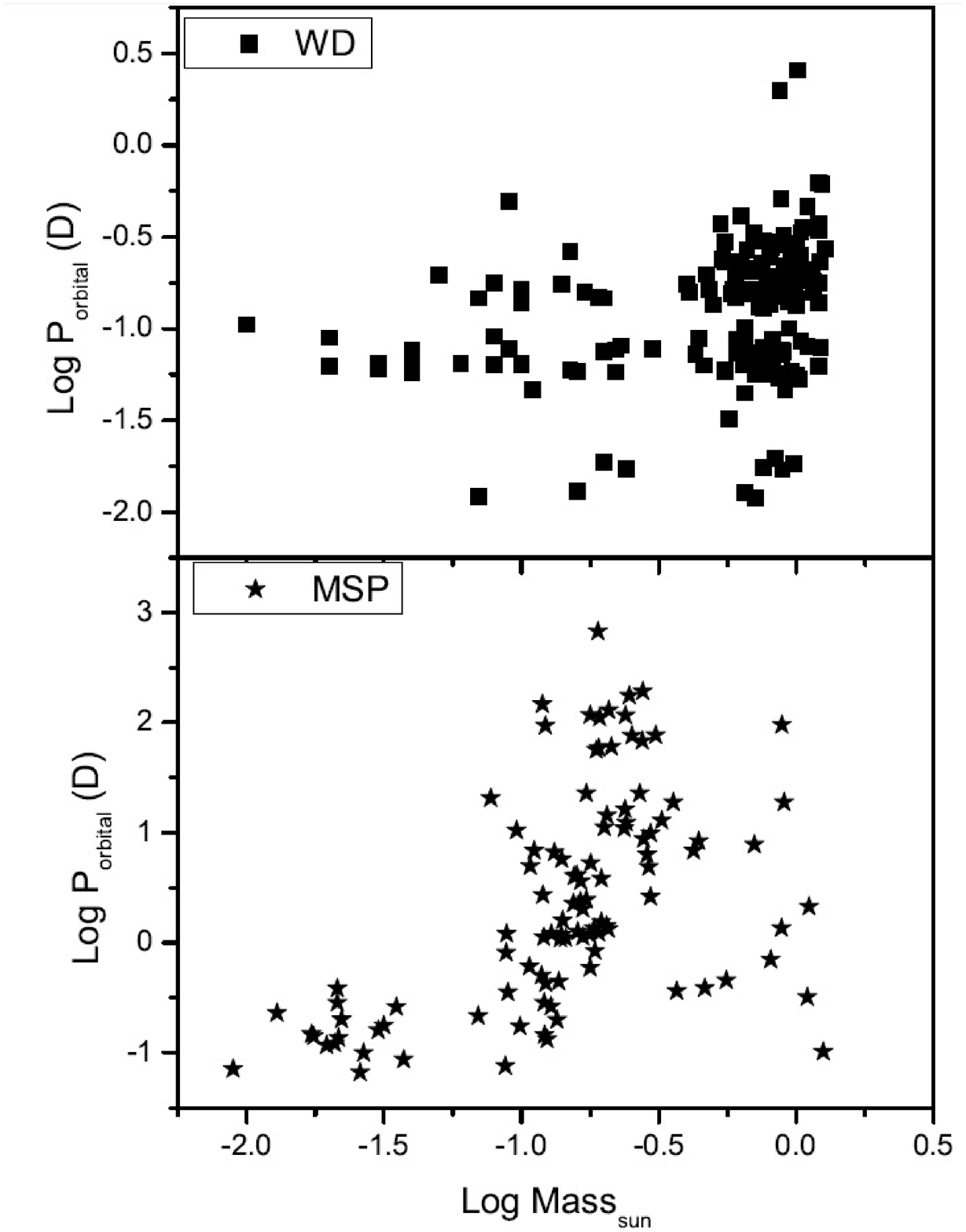}
\caption{The relation between masses and orbital periods,
showing that massive WDs recycle to MSPs. The orbital
periods  show a clear difference (over two
orders of magnitude). It should be noticed that the AIC process
leads to dynamical interactions in the systems during the conversion
of a WD into an MSP. The data for MSPs and WDs are taken from
the ATNF catalogue \citep{2005AJ....129.1993M} and from \citep{2011yCat....102018R}
respectively.} \label{WD-MSP}
\end{center}
\end{figure}

\section{The Evolutionary Path from WDs to MSPs}
Fig.~4 is a simple flowchart of the currently favored model that is used to
explain the formation of various types of the systems. The left
column illustrates a binary system in which a more massive star
has typically a mass in the range of $8-11M_{\odot}$ and produces a
WD after exhausting its available thermonuclear fuel.

Later, as the binary system sheds its angular momentum through the wind, the companion
eventually fills its Roche lobe and begins accreting onto the WD.
The accretion causes the mass of the WD to steadily increase until it
approaches the Chandrasekhar limit. Thereafter, the WD
becomes unstable and undergoes  an AIC,  forming an NS, or detonates
completely as a Type Ia SN in a thermonuclear explosion.  
Our major assumption is that the WD--MS channel for producing MSPs is more common than the WD-NS channel, because (1) not all WDs have the right composition (O, Ne, Mg), and not all of them accumulate a sufficient mass ($\rm \Delta m = 0.1 - 0.2 M_{\odot}$) to undergo the AIC \cite{2010ChPhL..27k9801W};  (2) in the WD--NS case the progenitor star of the NS is more massive and hence may bring the binary into the Roche lobe overflow from a larger original separation.

 \begin{figure}
 \begin{center}
\includegraphics[width=10.0cm, angle=0]{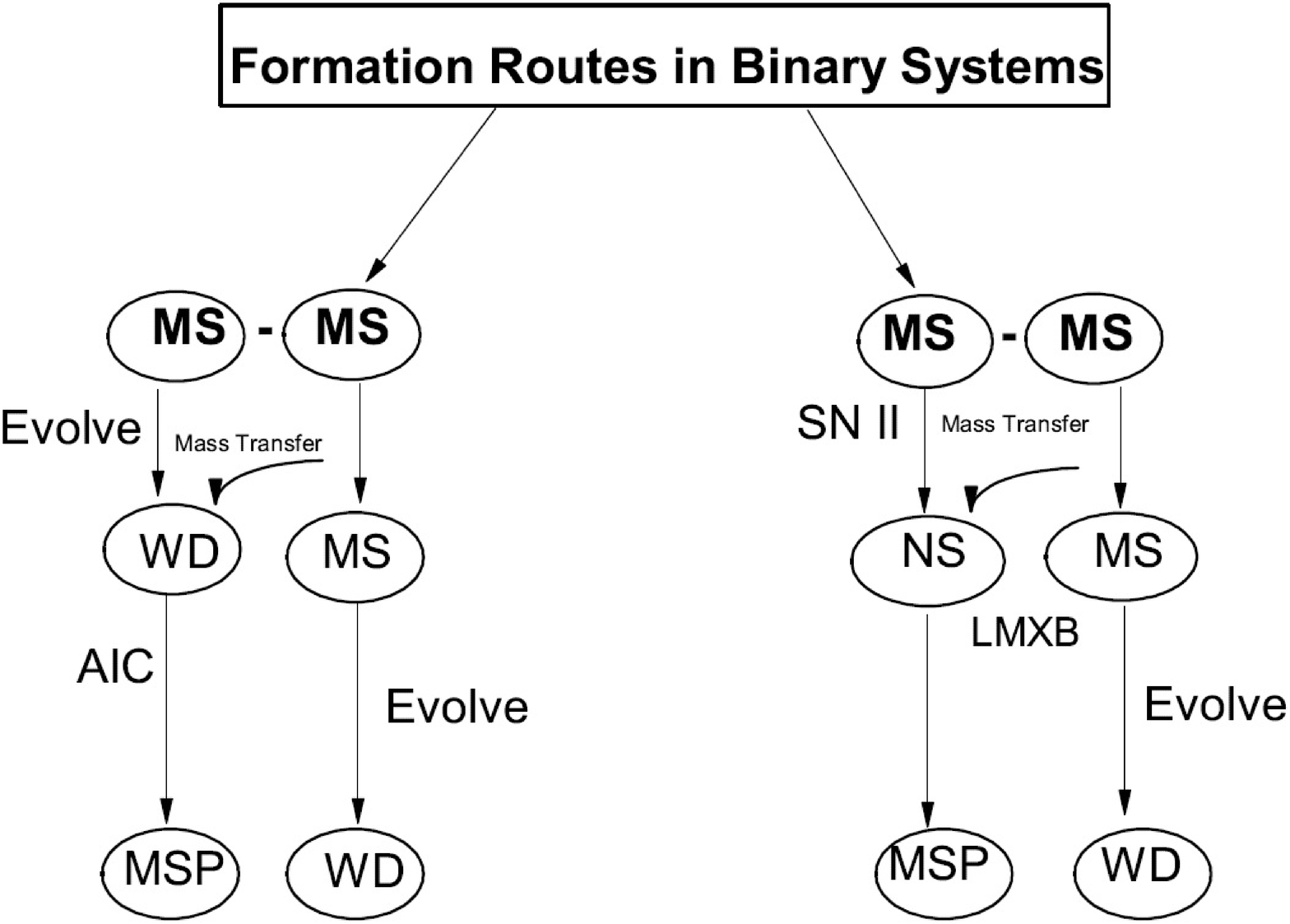}
\caption{A diagram illustrating possible binary evolution
scenarios leading to the formation of WD--MSP systems in the
Galactic disk.} \label{formation-1}
\end{center}
\end{figure}

In the right column of Fig. 4, the companion star is massive
enough so that it explodes directly as a Type II SN, producing an NS.
The binary does not stay intact if more than half the total
pre-supernova mass is ejected from the system during the
explosion \cite{2009ASSL..357....1L, 2020ApJ...903...88M}. 
After a time of $10^{7} - 10^{8}$~yrs, the old
spun-down NS can gain a new lease on life as a pulsar by accreting
matter and angular momentum at the expense of the orbital angular
momentum of the binary system. During this accretion phase, X-rays
are emitted by the frictional heating of the matter infalling to the
NS. This makes the system visible as an X-ray binary. At the end
of this spin-up phase, the secondary sheds its outer layers to
become a WD in an orbit around a now rapidly spinning MSP.

\section{Conclusions}

I have attempted to follow the AIC scenario channel in the considered evolutionary stage of close binary MSPs since some  circular binaries are found both in the Galactic disk and in globular clusters.
I present and analyze a whole set of terminal evolution tracks from WDs to the MSP stage that are suitable for the study of their formation and evolution including the mass exchange phases. With a focus on the combination of stellar parameters such as mass, P$_{orb}$, mass accretion rate, and eccentricity, I argue that our data appear to be more consistent with this scenario. However, knowledge of these tracks is required to study the cosmological perspective and any effects on galaxy evolution models. In addition, the existence of MSPs, WDs, and SNe Ia has
expanded the possible evolutionary channels for the formation
of near-Chandrasekhar-mass models
and MSPs appearing via the AIC process. 
In forthcoming work, I will expand on this research with studies aimed at delving deeper into these issues.


\bibliographystyle{JHEP}

\bibliography{nls.bib}





\end{document}